# Exploring the Charge Localization and Band Gap Opening of Borophene: A First-Principles Study

Andrey A. Kistanov,[a, b] Yongqing Cai,*[b] Kun Zhou,*[a] Narasimalu Srikanth,[c] Sergey V. Dmitriev[d,e] and Yong-Wei Zhang*[b]

Recently synthesized two-dimensional (2D) boron, borophene, exhibits a novel metallic behavior rooted in the s-p orbital hybridization, distinctively different from other 2D materials such as sulfides/selenides and semi-metallic graphene. This unique feature of borophene implies new routes for charge delocalization and band gap opening. Herein, using first-principles calculations, we explore the routes to localize the carriers and open the band gap of borophene via chemical functionalization, ribbon construction, and defect engineering. The metallicity of borophene is found to be remarkably robust against H- and F-functionalization and the presence of vacancies. Interestingly, a strong odd-even oscillation of the electronic structure with width is revealed for H-functionalized borophene nanoribbons, while an ultra-high work function (~ 7.83 eV) is found for the F-functionalized borophene due to its strong charge transfer to the atomic adsorbates.

**1. Introduction**

Recent success in synthesizing atomically thin two-dimensional (2D) borophene on Ag (111) substrates[1,2] has stimulated great interest in exploring the growth, structure and properties of this elemental 2D material.[3-7] As a magic element with coexistence of covalent and ionic characters, boron can show a versatile electronic structure, including semiconducting, semi-metallic, and metallic phases.[8-10] Previous studies showed that free-standing borophene exhibits a highly anisotropic electronic structure.[11,12] With its high carrier concentration at the Fermi level, which is absent in graphene with a zero carrier density at the Dirac cone, the atomically thin borophene serves as an ideal platform for investigating the distribution and response of electron gas confined in an ultrathin layer with external perturbations.

So far, various atomic models for boron clusters and 2D models have been theoretically proposed.[13-15] Concerning the structure of borophene, recent experiments reported two dramatically different atomic structures on the Ag substrate: the closely packed structure[1] and the hole-containing structure.[2] While experiments claimed the stability of a borophene sheet supported by the Ag substrate, the stability of a free-standing borophene is still elusive. First-principles phononic calculations showed that the long-wavelength imaginary modes always exist in isolated triangular sheets and other polymorphs like β$_{12}$ sheets (refer to phonon dispersion in the supplementary information in Ref. 14). Physically or chemically bound species were found to promote the stability of borophene.[16,17] Recent theoretical studies proposed a viable approach of using substrates[18,19] or chemical functionalization[20] to tune the adsorption and energetics of borophene. In principle, a proper truncation of the borophene sheet by breaking the lattice periodicity can eliminate these long-wavelength softening phonons, thus stabilizing structures without the need of chemical functionalization.

As the boron atom has three valence electrons, it needs to pair with five additional electrons to satisfy the octet rule. However, in borophene, each boron atom forms bonds with six neighbors, thus favoring the metallic phase according to the band theory. For electronic applications of borophene, an intriguing issue is the localization of its itinerant electrons and ultimately its band gap opening. Approaches for the band gap opening in 2D materials can be categorized into two groups: i) quantum confinement induced by the construction of finite-sized structures, like ribbons, edges and dots,[21-23] and ii) chemical functionalization.[24,25] However, the effectiveness of both approaches on the borophene band gap opening remains unclear. The formation of ionic bonds in a high-pressure boron phase[26] suggests a different charge distribution in boron materials in comparison with other 2D materials, especially graphene. The failure of the octet rule in pure borophene implies a new mechanism of the charge localization/delocalization, which is yet to be understood.

In this work, we explore the routes for the charge localization and the band gap opening of borophene via chemical functionalization, defect engineering, and ribbon construction by using first-principles calculations. We show that the metallicity of borophene is remarkably robust against H- and F-functionalization and the presence of vacancies. Interestingly, a strong odd-even oscillation of the electronic structure with the ribbon width is revealed for H-functionalized borophene nanoribbons, and band gap opening occurs only for a specific type of ribbons. Owing to the high density of states (DOS) near the Fermi level, a record-high work function is found for the F-functionalized borophene. These unique features of chemically functionalized borophene sheets and ribbons may indicate many interesting applications.

## 2. Methods

Our theoretical calculations are performed by using the Vienna *ab initio* simulation package (VASP)[27] within the framework of the density functional theory (DFT). The generalized gradient approximation (GGA) with the Perdew-Burke-Ernzerhof (PBE) functional is selected together with an energy cutoff of 450 eV. The effects of spin polarization are considered in our calculations. For various chemically modified 2D borophene, a 20×15×1 Monkhorst-Pack grid is used for the k-point sampling in the first Brillouin zone. For the line (zigzag)-edge borophene nanoribbons (BNRs), a 20×1×1 (1× 8×1) k-point sampling is used. To avoid the spurious interaction between periodical images, a vacuum space of 15 Å along the out-of-plane direction is created. All the structures are relaxed until the forces become smaller than 0.01 eV/Å. To more accurately describe the self-interaction and screening of carriers, the hybrid functional Heyd-Scuseria-Ernzerhof (HSE06)[28] is adopted for the work function calculations.

## 3. Results and discussion

We first explore the modification of the electronic properties via functionalization of borophene surface by H, F, and O atoms. We notice that several atomic models, showing long-wavelength negative phonon modes, have been proposed for monolayer borophene to match the experimental images of the samples grown under different conditions.[14] In this work, the closely-packed atomic model, as revealed in a recent experiment,[1] is chosen (see Figure 1a) since other atomic models are its direct derivatives. The considered closely-packed model, which consists of equilateral triangles as the basic unit, has a *Pmmn* space group with a rectangular unit cell. It can be regarded as a staggered honeycomb lattice with additional atoms located at the hexagon centers. Each rectangular unit cell contains two symmetrically inequivalent boron atoms occupying the two 1a Wyckoff sites at the corner and face center positions, respectively. The relaxed lattice constants are 1.618 and 2.864 Å along the *a* (line-edge) and *b* (zigzag) directions, respectively. To examine the extreme effect of the chemical functionalization, we consider a 50% coverage of borophene surface with H- and F-functionalizing groups above a 2×2 supercell due to the balance for maximizing the doping effect while maintaining the stability of the doped structure. Considering the relatively small radius of the boron atom, this concentration is quite high.

Figures 1b and c show the optimized structures and the electronic band structures of pristine, H-, and F-functionalized borophene sheets with adsorption on both sides. Consistent with previous works,[29,30] the pristine borophene shows a metallic behavior (see Figure 1a) with the Fermi level crossing the electronic levels. Owing to the puckered zigzag structure (side view), the pristine borophene shows an anisotropic electronic structure with half-filled bands along the Γ-X (line-edge) direction but an energy gap along the Y-Γ (zigzag) direction. Such an intriguing electronic property implies an orientationally different quantum confinement effects in borophene, which may lead to angle-dependent plasmonic behavior in this ultrathin metallic sheet.

Compared with the pristine borophene, the H-functionalized borophene only shows a small change in its lattice constant. In contrast, an F-functionalized borophene sheet shows significantly deformed B-B bonds, leading to a large distortion of the host borophene lattice. Such difference can be attributed to the strong subtraction of electrons from the sheet to the anionic adsorbents, in comparison with the H- group. Interestingly, the F-functionalization of borophene induces changes in the lattice constant along the zigzag direction, around 1%, while the change along the line-edge direction is only slight. Concerning the electronic properties, surprisingly, all these highly chemically functionalized borophene sheets remain metallic. This is in strong contrast to graphene where hydrogenation and fluorination are well known to lead to band gap openings. Therefore, the charge localization, which is necessary for the band gap opening, is hard to induce via the surface chemical functionalization of borophene. Projected band analysis for the H-functionalized borophene (see Figure 1b) shows that the states around the Fermi level are predominantly populated with the H states. In contrast, the F states in the F-functionalized borophene are largely distributed below the Fermi level (see Figures 1c). In addition, upon the functionalization by these atoms, partially occupied levels are formed along the Y-Γ direction, which is otherwise empty in pristine borophene. Therefore, selective atomic functionalization enables the modulation of the anisotropy in the electronic properties of borophene. It is noted that the in-plane doping within the atomic sheet of borophene is able to open the band gap.[31]

The above-mentioned robust metallicity in these surface functionalized borophene sheets is absent in graphene and transition metal dichalcogenides (TMDs). The finite density of these free carriers at the Fermi level suggests that borophene and its functionalized derivatives are promising for applications as interconnecting and field-emitting materials. Since the work function, which quantifies the ability of electrons to move from the surface of a material to vacuum, is critically important for field emission and rectification of conducting barriers,[32,33] in the following, we examine the change in the work functions of these functionalized borophene sheets.

Figure 1d shows the energetic diagram of the work functions for various functionalized borophene sheets in comparison with other common bulk metals and graphene. From the diagram, the following important features can be identified. Firstly, the work function of a pristine borophene sheet is 5.31 eV (obtained via HSE calculation), which is larger than that of most listed metals, except Pt. Moreover, the work function of pristine borophene is also higher than that of graphene (~ 4.5 eV).[34] This is surprising since a carbon atom has a larger electronegativity than a boron atom. The higher work function of borophene could be attributed to the nature of atomic states around the Fermi level. Borophene mainly consists of in-plane *s-p* hybridized (σ) states, which are lying lower than those of the out-of-plane $p_z$ (π) states in the graphene case. Thus, an electron in borophene is harder to knock out than that in graphene. Secondly, the work function of borophene increases slightly to 5.88 eV for the H-functionalized and dramatically to 7.83 eV for the F-functionalized borophene. The underlying origin may arise from the strong dipole layer pointing inward towards the central borophene layer due to its transfer of electrons to the functionalizing atoms (see Figure 1d). In other 2D materials, like graphene and TMDs, the density of electrons is negligible at the Fermi level, which means that the magnitude of the dipole layer is modest upon the chemical functionalization due to the limited charge transfer. In contrast, a borophene layer has a considerably high density of carriers at the Fermi level because of its intrinsic metallicity, giving rise to a pronounced charge flow and a built-in dipole layer. This great tunability in the work function suggests that the chemically functionalized borophene sheets can be used as a buffer layer for reducing the contacting resistance and Schottky barrier at the interface. In addition, the high work function in the F-functionalized borophene sheet is particularly useful for electron collection and hole injection.

Next, we investigate the effect of the atomic vacancies on the electronic properties of borophene (see Figure 2). We consider both monovacancies (MV) and divacancies (DV) with the loss of one and two boron atoms in the 6×5 supercell (56 atoms), respectively. In perfect borophene, each boron atom has a coordination number of six. With the creation of an MV, six peripheral atoms become fivefold coordinated and the defect core has a local symmetry of $C_{2v}$ (see Figure 2b). For the DV, two different configurations exist: horizontal ($C_{2v}$ symmetry) and tilted ($C_i$ symmetry) DV, depending on the relative direction of the deleted boron dimer to the $a$ lattice (see Figures 2c and d). The horizontal DV is slightly more stable with the energy of 0.07 eV lower than that of the tilted DV. In the DV structures, there are two new-born fourfold boron atoms in the edge in addition to the six fivefold corner atoms. The formation energy ($E_{form}$), which is defined as $E_{form}=E_{perfect}-E_{defect}-nE_{atom}$, where $E_{perfect}$, $E_{defect}$ and $E_{atom}$ are the total energies of the perfect and defective borophene, and the single boron atom, respectively, and $n$ is the total number of the removed atoms. The calculated values of $E_{form}$ for MV, horizontal, and tilted DVs are -5.61, -12.58, and -12.51 eV, respectively. The relatively high values of $E_{form}$ suggest that it is difficult to form isolated vacancies in borophene. However, the energy cost for forming an MV in borophene is smaller than that in graphene (-7.57 eV) and that of a boron MV in h-BN (from -7.50 to -10.20 eV).[36]

It is well known that the electronic properties around the vacancy core may change dramatically owing to the breaking of the lattice periodicity.[37-39] Figure 2a shows the DOS in the perfect, MV-, and DV-containing borophene sheets. It is seen that the metallicity of these borophene sheets is robust against the presence of vacancies. Interestingly, the Fermi level significantly shifts upwards for the MV and DV cases compared with perfect borophene (see the arrows in Figure 2a). The bands of vacancy-containing borophene sheets are non-zero at the Fermi level and mainly contributed by the $p_z$ orbitals of B atoms. Different from new localized states formed in the band gap-associated vacancies in MoS$_2$[40] and phosphorene[41] cases, there are no peaks related to the dangling bond states due to vacancies in borophene. However, by comparing the local DOS (LDOS) of peripheral atoms in the defect core and those of atoms far from the vacancy center (see Figures 2b-d, bottom panel), we find that there is a significant difference in the LDOS profiles, suggesting that the states are renormalized greatly in the defect center.

To explore the routes to introduce localized states in the intrinsic metallic borophene, we examine one-dimensional (1D) nanostructures of borophene, that is, nanoribbons. Both the line- and zigzag-edge ribbons are considered. Following the normal nomenclature of nanoribbons in graphene, MoS$_2$, and phosphorene, the BNR along the line-edge or zigzag direction is named as LE-$N_d$ BNR or ZZ-$N_z$ BNR according to the number of B-B dimer lines ($N_d$) or zigzag chains ($N_z$) across the ribbon width. Figures 3a and b show the atomic models and the band structures of pristine LE-9 and LE-10 BNR, selected as the representatives for the odd- and even- width BNRs, respectively. It is seen that both types of BNRs are metallic without any localized states and band gap opening, which is in a strong contrast with graphene.[42] The orbital-resolved band structures suggest that the out-of-plane $p_z$ and in-plane $p_y$ orbitals of B atoms are dominant at the Fermi level, which may account for the quasi-planar structures at the edges (red circles in Figures 3a and b). The transporting states consisting of $p_x$ orbitals (blue lines in Figures 3a and b) and aligned along the momentum (Γ-X) direction are quantized with an energy gap in the band dispersion. The isosurface plots of electronic densities of the valence band clearly reflect this quantized feature of $p_x$ states with regular nodal planes across the width direction of ribbons. The formed quasi-1D $p_x$ states are highly delocalized, which may facilitate the stabilization of the LE edge-terminated BNRs. In contrast, pristine ZZ BNRs undergo severe structural distortions and become disordered after structural relaxation (not shown).

Interestingly, the structural integrity is largely maintained for both H-functionalized LE (Figure 3c) and ZZ (Figure 3d) BNRs, which are created through cutting the 2D H-functionalized borophene, as shown in Figure 1b. The plots of the band structure show an odd-even oscillation of the band gap with the ribbon width for both H-functionalized LE and ZZ BNRs: the band gap is absent for odd H-functionalized BNRs but present for even H-functionalized BNRs. For instance, the band gap is 0.93 eV for LE-8 H-BNR while zero for LE-9 H-BNR. The band gap is 1.07 eV for ZZ-16 H-BNR but zero for ZZ-15 H-BNR. Therefore, depending on the width of the H-functionalized BNRs, the electronic states can become partially localized and have a band gap opening, as supported by the isosurface plots of the partial charge density of the valence bands in Figure 3. The orbital-resolved band structure plots (see Figure 3c) show that the frontier orbitals in LE H-BNRs are still $p_y$ and $p_z$ orbitals, similar to those in the pristine BNRs (see Figures 3a and b). For the ZZ H-BNRs, the frontier orbitals mainly consist of $p_x$ components, as shown by the blue dispersion lines. By examining the charge density distribution of the LE-8 H-BNRs (Figure 3c), we find that such band gap opening in the specific type of H-BNRs is due to hydrogen-induced Peierls instability of the metallic states in the BNRs. This scenario is evidenced by the strong structural distortion of the BNR lattice and the tilted H-B bonds.

We also investigate the F-functionalized BNRs and find that their electronic properties are insensitive to the ribbon width. Therefore, only one ribbon is selected as a representative for each of LE and ZZ BNRs. Figures 4a and b show the optimized atomic structure and the band structure of F-functionalized LE-10 and ZZ-16 BNRs, respectively. The orbital-resolved band structures show that F-BNRs remain metallic, the same as pristine BNRs. Different from the graphene case, where fluorination or ribbon construction can effectively open the band gap, the coexistence of fluorination and ribbon construction is not able to cause a metal-semiconductor transition, implying the robustness of metallicity in borophene. For the F-functionalized ZZ BNR, the in-plane $p_y$ orbital of B atoms is dominant at the Fermi level, which is different from the H-functionalized ZZ BNR with $p_x$ as the frontier orbitals (see Figure 3d), suggesting that these functional groups can have a selective hybridization of the boron orbitals and alter the orbital population at the Fermi level.

To examine the stability of the considered structures, we calculate the average binding energy ($E_b$) of perfect borophene which is defined as $E_b = (n_B E_B - E_{tot})/n$, and the $E_b$ of H-, F,- and O-functionalized borophene using $E_b = (n_B E_B + n_X E_X - E_{tot})/n$, where $E_{tot}$ is the total energy of the functionalized system, $E_B$ is the energy of a single boron atom, $E_X$ is the energy of a single H or F atoms, $n_B$ is the total number of B atoms, $n_x$ is the total number of H or F atoms, and $n$ is the total number of atoms in the system. The calculated values for the $E_b$ for perfect, H-, and F-functionalized borophene are 5.86, 4.78, and 5.25 eV, respectively. Clearly, all the three considered structures of the functionalized borophene show a better stability than the pristine borophene. To verify the stability of BNRs, $E_b$ is also calculated and shown in Table 1. It is seen that the values of $E_b$ are positive for both LE-BNRs and ZZ-BNRs, indicating that these nanoribbon derivatives are energetically stable, which is in a good agreement with the recent work.[43] To confirm the calculation results of $E_b$, we also consider the edge energy ($E_{edge}$), which is defined as $E_{edge} = (E_{BNR} - nE_B)/L$, where $E_{BNR}$ is the total energy of the BNR, $L$ is the length of the ribbon along the periodic direction, $E_B$ is the total energy per atom in 2D borophene, and $n$ is the total number of atoms. As shown in Table 1, the calculated values of $E_{edge}$ have the following sequence: 0 < LE BNRs < LE H-BNRs < LE F-BNRs and 0 < ZZ F-BNRs < ZZ H-BNRs, confirming the stability of BNRs.

In addition, we also perform *ab initio* molecular dynamics (AIMD) calculations at 300 K for 8 ps using the Nose-Hoover method to check the stability of the functionalized and vacancy-containing borophene, as well as the pristine and functionalized BNRs. The snapshots of the simulation results are shown in Figure 5. It is seen that during this long time (in terms of *ab initio* calculations), all the considered structures are stable. It should be noted that the ZZ H- and F-BNRs (Figures 5g and i) exhibit a lower stability than the LE H- and F- BNRs.

**Table 1.** The average binding energies ($E_b$) and edge energies ($E_{edge}$) for the borophene nanoribbons.

| BNRs type | $E_b$, eV | $E_{edge}$, eV/Å |
|---|---|---|
| LE-9 BNRs | 5.89 | 0.062 |
| LE-10 BNRs | 5.90 | 0.066 |
| LE-8 H-BNRs | 4.85 | 2.170 |
| LE-9 H-BNRs | 4.84 | 2.090 |
| ZZ-15 H-BNRs | 4.80 | 4.330 |
| ZZ-16 H-BNRs | 4.80 | 4.350 |
| LE-9 F-BNRs | 5.49 | 2.600 |
| LE-10 F-BNRs | 5.51 | 2.670 |
| ZZ-15 F-BNRs | 5.52 | 4.040 |
| ZZ-16 F-BNRs | 5.52 | 4.190 |

Currently, it is still a challenge to obtain free-standing borophene experimentally, and thus double-side functionalized borophene. Nevertheless, various attempts are being made to tackle this issue. One possible way to address this challenge is to grow a borophene layer on a substrate with a weaker interaction, and then the attachment and the transfer of the borophene layer may be achieved according to the previous theoretical prediction.[18] The other way is via surface decoration of supported borophene by creating new bonding states at the surface and edge boron atoms. In this case, dopant atoms may diffuse from the edges to the interior of the borophene-substrate interface, allowing for possible exfoliation and double-side functionalization of borophene.

In this work, we identify possible avenues for the band gap opening and the charge localization through creating vacancies, forming edges, and surface functionalization. These methods are found to be effective in the band gap adjustment in other 2D materials. However, according to our present simulation, there is no band gap opening from vacancies and chemical functionalization of the pristine 2D borophene, signifying its robust metallicity.

To further confirm the absence of a localized state with the surface functionalization and local vacancies, we calculate the electronic localization functions (ELFs) of various borophene derivatives and the results are shown in Figure 6. The value of the ELF (between 0 and 1) reflects the degree of the charge localization in the real space, where 0 represents a free electronic state while 1 represents a perfect localization. The isosurface value of 0.65 is adopted in Figure 6. For the case of vacancies in borophene (Figures 6a-c), it can be seen that a significant number of electrons are distributed between the B-B bonds along the [001] direction both around and far from the vacancy, indicating a strong covalent chemical hybridization. There exists a zero ELF along the [110] and [110] directions, indicating highly delocalized electronic states. This is different from other semiconducting 2D materials, like $MoS_2$, where defective states are only localized at the defect cores. A similar situation occurs for the cases of the H- and F-dopants and H-BRNs (Figures 6d-g), where the ELF is zero at the boron atoms, implying that the electrons around the boron atoms are highly delocalized.

**Conclusions**

We have investigated the electronic properties of surface functionalized borophene sheets and explored the possible avenues for opening the band gap of borophene via systematic first-principles calculations. We have found that the metallicity in borophene is immune to the surface functionalization and the presence of vacancies. Interestingly, the anisotropy of the electronic properties and the nature of the orbitals at the Fermi level can be altered upon the surface functionalization, enabling the modulation of the borophene properties. Due to the high density of itinerant electrons in the atomically thin borophene sheet, the band gap opening via quantum confinement, which is effective for graphene, becomes ineffective for borophene. We have also revealed that the work function of borophene can be tuned to a large degree as the high electronic gas confined in the atomically thin sheet of borophene can be transferred to the functionalizing groups.

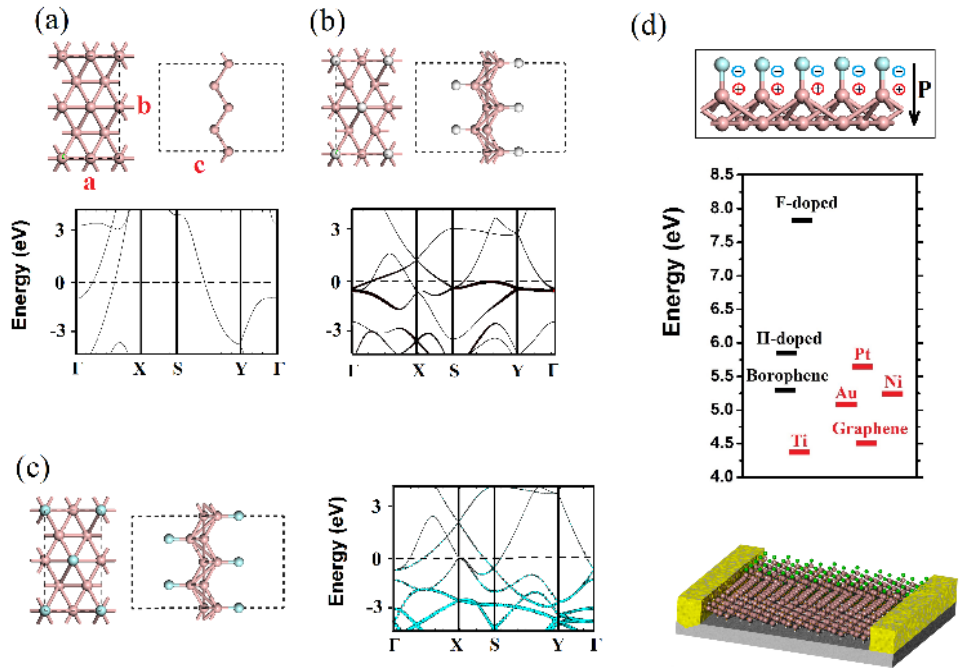

**Figure 1.** The atomic configuration and the band structure of (a) pristine, (b) H-, and (c) F-functionalized borophene. The component of the states scales with the radius of the black and cyan circles for H- and F-functionalized cases, respectively. (d) Inward dipole layer built in F-functionalized borophene due to the charge transfer (top panel). Comparison of the work functions of pristine, H-, and F-functionalized borophene (calculated with HSE method) with those of the common metals and graphene (middle panel). Schematic plot of the integration of the chemically functionalized borophene for improving the efficiency of the injection and transport of carriers in nanoelectronics devices (bottom panel). The work functions of graphene and other metals are adopted from Refs. 34 and 35.

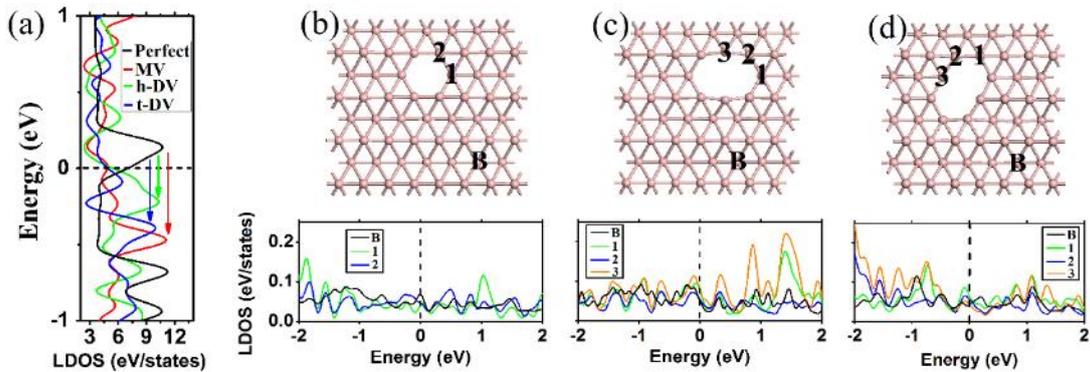

**Figure 2.** (a) Comparison the total DOS of perfect (black line), MV-containing (red line), horizontal DV-containing (green line), and tilted DV-containing (blue line) borophene. The atomic structure (upper panel) and LDOS (bottom panel) for several edge atoms of (b) the MV-containing, (c) the horizontal DV-containing, and (d) the tilted DV-containing borophene sheets. The LDOS of the labeled edge atoms (atoms 1, 2, and 3 considered due to the symmetry reason) are compared with that of the boron atom (black line) far from the core of the vacancy. Black dashed lines show the position of the Fermi level.

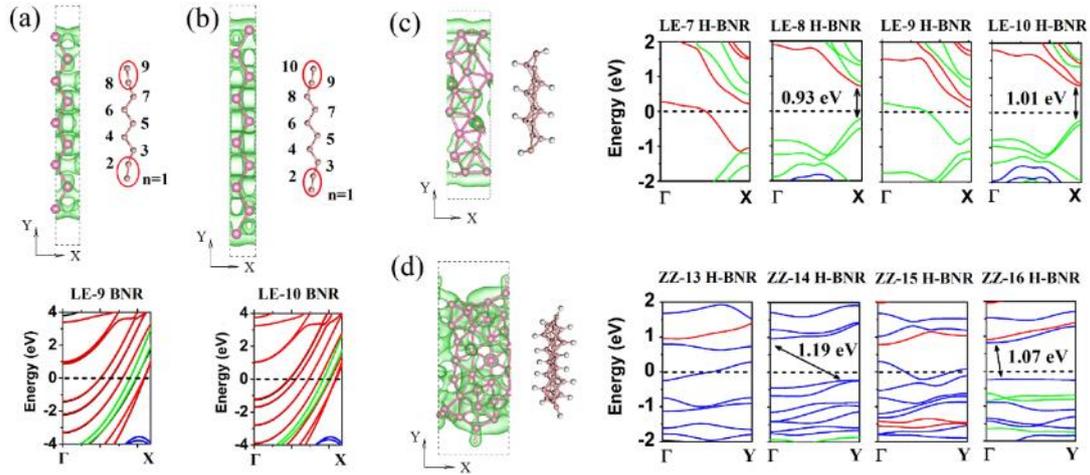

**Figure 3.** The Atomic model and the band structure of pristine (a) LE-9 and (b) LE-10 BNR. The top and side views of the atomic configurations, together with the charge distribution at the top of the valence band, are plotted in the upper panel. The atomic models of LE-8 and ZZ-15 H-BNRs with charge distribution for the states from $E_f$-1eV to $E_f$ are shown in the right panels of (c) and (d), respectively. The width-dependent odd-even oscillation of the band gap opening in LE and ZZ H-BNRs is shown in the left panels of (c) and (d), respectively. The blue, green, and red curves on the band structure plots represent the projected states for $p_x$, $p_y$, and $p_z$ orbitals of B atoms, respectively. The black dashed lines show the Fermi level.

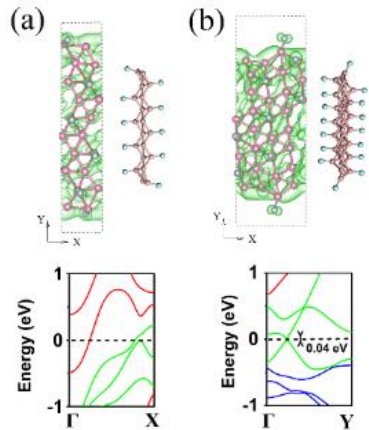

**Figure 4.** (a) The atomic configuration and (b) the band structure of the F-functionalized BNRs. Integrated charge densities from $E_f$-1 eV to $E_f$ are shown in (a, upper panel) LE and (b, upper panel) ZZ ribbons. The blue, green, and red curves in the band structure plots represent the occupation of the $p_x$, $p_y$, and $p_z$ orbitals of B atoms.

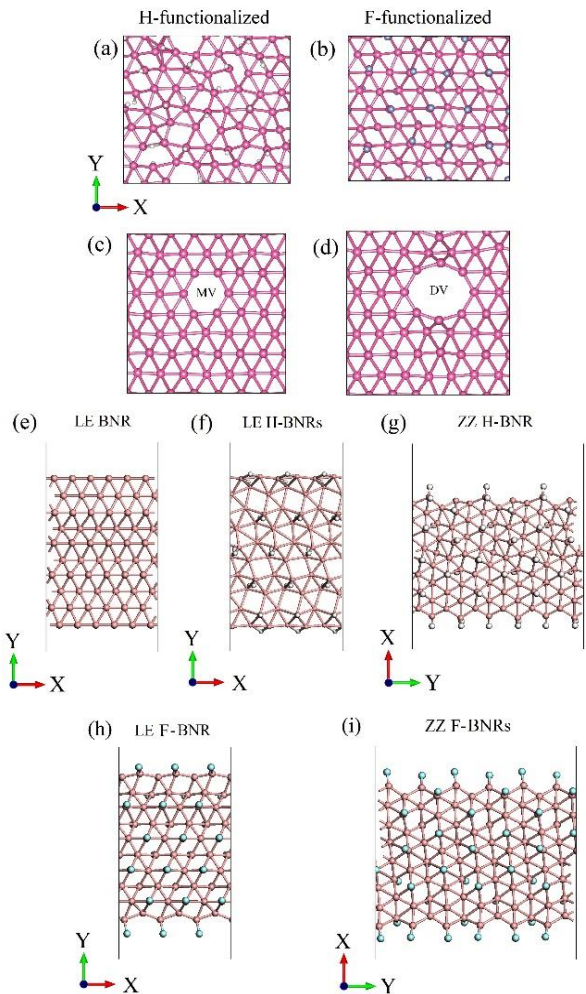

**Figure 5.** Snapshots of the (a) H- and (b) F-functionalized borophene, (c) MV- and (d) DV-containing borophene, (e) LE, (f) LE H-, (g) ZZ H-, (h) LE F- and (i) ZZ F-BNRs calculated by AIMD at 300 K. Atoms B, H, and F are colored in pink, white, and purple.

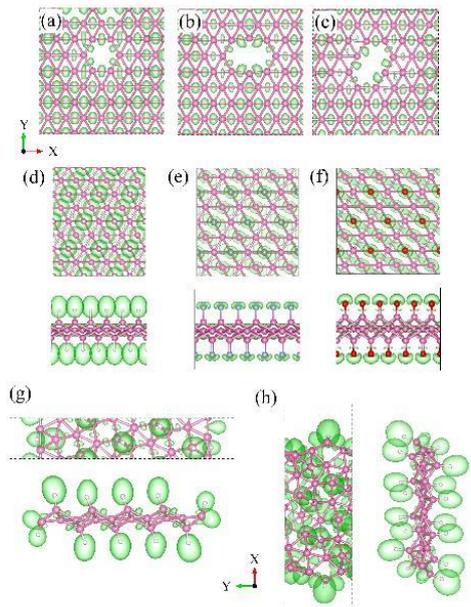

**Figure 6.** The ELFs for (a) MV-, (b) horizontal DV-, and (c) tilted DV-containing phosphorene sheets, (d) H- and (e) F-functionalized borophene, and (f) LE-10 and (g) ZZ-16 H-BNRs.